\documentclass[iop]{emulateapj}
\usepackage{graphicx}
\usepackage{amssymb}
\usepackage{amsmath}
\usepackage{hyperref}
\usepackage{natbib}
\usepackage{longtable}
\usepackage{color}
\usepackage{float}
\usepackage[percent]{overpic}

\newcommand{\nar}{New Astron. Rev.}

\newcommand{\Kepler}{{\it Kepler}}

\newcommand{\kms}{km~s$^{-1}$}
\newcommand{\teff}{$T_\text{eff}$}
\newcommand{\jktebop}{{\sc jktebop}}

\newcommand{\msun}{M$_\odot$}
\newcommand{\rsun}{R$_\odot$}

\newcommand{\loglsun}{$\log{(L/L_\odot)}$}
\newcommand{\mjup}{M$_\text{Jup}$}

\newcommand{\sbratio}{0.0602}
\newcommand{\esbratio}{0.0024}
\newcommand{\sumradii}{0.0590}
\newcommand{\esumradii}{0.0018}
\newcommand{\ratradii}{0.268}
\newcommand{\eratradii}{0.025}
\newcommand{\inclination}{87.69}
\newcommand{\einclination}{0.14}

\newcommand{\period}{7.0504829}
\newcommand{\eperiod}{0.0000047}
\newcommand{\tzero}{2456916.65777}
\newcommand{\etzero}{0.00014}
\newcommand{\rvprim}{19.64}
\newcommand{\ervprim}{0.11}
\newcommand{\rvsys}{5.695}
\newcommand{\ervsys}{0.084}

\shorttitle{An Eclipsing Binary in the Pleiades}
\shortauthors{David et al.}

\begin{document}

\title{HII 2407: A Low-Mass Eclipsing Binary Revealed by K2 Observations of the Pleiades}

\author{Trevor J. David \altaffilmark{1,2}, John Stauffer \altaffilmark{3}, Lynne A. Hillenbrand \altaffilmark{1}, Ann Marie Cody \altaffilmark{4}, Kyle Conroy \altaffilmark{5}, Keivan G.\ Stassun \altaffilmark{5,6}, Benjamin Pope \altaffilmark{7}, Suzanne Aigrain \altaffilmark{7}, Ed Gillen \altaffilmark{7}, Andrew Collier Cameron \altaffilmark{8}, David Barrado \altaffilmark{9}, L.M. Rebull \altaffilmark{3}, 
Howard Isaacson \altaffilmark{10}, Geoffrey W. Marcy \altaffilmark{10},
Celia Zhang\altaffilmark{1}, Reed L. Riddle\altaffilmark{1},  Carl Ziegler\altaffilmark{11}, Nicholas M. Law\altaffilmark{11}, Christoph Baranec\altaffilmark{12} 
}

\altaffiltext{1}{Department of Astronomy, California Institute of Technology, Pasadena, CA 91125, USA}
\altaffiltext{2}{NSF Graduate Research Fellow}
\altaffiltext{3}{Spitzer Science Center, California Institute of Technology, Pasadena, CA 91125, USA}
\altaffiltext{4}{NASA Ames Research Center, Mountain View, CA 94035, USA}
\altaffiltext{5}{Department of Physics \& Astronomy, Vanderbilt University, Nashville, TN 37235, USA}
\altaffiltext{6}{Department of Physics, Fisk University, Nashville, TN 37208, USA.}
\altaffiltext{7}{Department of Physics, University of Oxford, Keble Road, Oxford OX1 3RH, UK}
\altaffiltext{8}{SUPA, School of Physics and Astronomy, University of St Andrews, North Haugh, St Andrews, Fife KY16 9SS, UK}
\altaffiltext{9}{Centro de Astrobiolog\'{i}a, INTA-CSIC, Dpto.\ Astrof\'{i}sica, ESAC Campus, P.O.~Box 78, 28691 Villanueva de la Ca\~nada, Madrid, Spain}
\altaffiltext{10}{Department of Astronomy, University of California, Berkeley CA 94720, USA}
\altaffiltext{11}{Department of Physics and Astronomy, University of North Carolina at Chapel Hill, Chapel Hill, NC 27599-3255, USA}
\altaffiltext{12}{Institute for Astronomy, University of Hawai`i at M\={a}noa, Hilo, HI 96720-2700, USA}

\email{tjd@astro.caltech.edu}

\begin{abstract}
The star HII 2407 is a member of the relatively young Pleiades star cluster and was previously discovered to be a single-lined spectroscopic binary.  It is newly identified here within \Kepler/$K2$ photometric time series data as an eclipsing binary system.  Mutual fitting of the radial velocity and photometric data leads to an orbital solution and constraints on fundamental stellar parameters. While the primary has arrived on the main sequence, the secondary is still pre-main-sequence and we compare our results for the $M/M_\odot$ and $R/R_\odot$ values with stellar evolutionary models. We also demonstrate that the system is likely to be tidally synchronized. Follow-up infrared spectroscopy is likely to reveal the lines of the secondary, allowing for dynamically measured masses and elevating the system to benchmark eclipsing binary status.
\end{abstract}


\section{Introduction}
\label{sec:introduction}
Binary stars, notably double-lined eclipsing binaries, are fundamental astrophysical systems whose study is key to obtaining accurate empirical measurements of stellar radii, masses, and temperatures. These precisely derived quantities are necessary for calibrating theoretical models of stars, and understanding stellar evolution.  Particularly valuable are well-characterized systems in either the pre-main sequence or post-main sequence phases where stellar evolution is more rapid, and fundamental calibrators correspondingly more rare relative to the main sequence. Among pre-main-sequence stars, fewer than 10 systems with masses below 1.5 M$_\odot$ have published orbital solutions and fundamentally derived stellar parameters; see \cite{stassun2014} and \cite{ismailov2014} for reviews.

The Pleiades cluster (d = $136.2 \pm 1.2$ pc; \cite{melis2014} and age = $125 \pm 8$ Myr; \cite{stauffer1998})   is well-studied and has sizable membership.  At the upper end of the mass distribution the stars are slightly evolved, with a well-populated main sequence between $\sim$0.5 and $\sim$3 $M_\odot$ (or M0 through B8 spectral types), and at lower masses the stars are still contracting as pre-main sequence objects. The $K2$ phase of the Kepler mission \citep{howell2014} has observed $\sim$800 bona fide and candidate Pleiads.

We report here the detection of Pleiades member HII 2407 as an eclipsing binary system, and make use of $K2$ photometry and existing radial velocity measurements from the literature to derive an orbital solution and constrain the stellar parameters.  Mutual fitting of the data combined with assumptions based on available information about the early-K type primary suggests a mid-M type secondary. At the Pleiades age, the primary has arrived on the main sequence while the secondary is still pre-main-sequence. Future observations will be needed in order to detect the spectrum of the secondary distinctly from that of the primary, rendering it a double-lined system and enabling a unique solution for the masses of the individual components.

\section{K2 Observations and Analysis}
\label{sec:ktwo}

The observations took place during $K2$ Campaign 4 which ran from 2015-02-08 through 2015-04-20 UTC\footnote{Data release notes available at \url{http://keplerscience.arc.nasa.gov/K2/C4drn.shtml}}. Although we also produced our own light curve from aperture photometry, in our final analysis, we use the Simple Aperture Photometry (SAP) light curve available from the Mikulski Archive for Space Telescopes (MAST), which we corrected for systematics and intrinsic stellar variability.

The dominant characteristic of the $K2$ light curve of HII 2407 is a variability pattern of $\sim$2\% amplitude caused by rotational modulation of star spots. Primary eclipses of $\sim$5\% depth were detected by inspection of the raw light curve. Following removal of the spot modulation pattern, a Lomb-Scargle periodogram analysis of the corrected light curve yielded an orbital period of 7.05 days. Phase-folding the corrected light curve on this period then revealed secondary eclipses of depth $<0.5$\%. The light curve extends over $\sim$71 days, with $\sim$30 minute cadence, yielding ten 2015 epochs.

In addition to the eclipses and star spot variability pattern, the light curve displays a saw-tooth-like pattern induced by the roll angle variations of the satellite. Following \cite{aigrain2015}, we model the spot- and roll- induced variations jointly, using a Gaussian process (GP) model with three components: a time-dependent term to represent the spot modulation, a term depending on the star's position on the CCD (as measured via the centroid) to represent the systematics, and a white noise term. This enables us to subtract the systematics and, where appropriate, the spot-induced variability in order to study the eclipses. 

A detailed description of GP regression applied to $K2$ light curve modelling is beyond the scope of this paper; we refer the interested reader to \cite{aigrain2015} and restrict ourselves here to the differences between the present analysis and that paper. The position component was two-dimensional, depending on the centroid $x$ and $y$, rather than on a 1-D estimate of the roll-angle variations. The time component was modelled as quasi-periodic to reflect the periodic but evolving nature of the spot-induced variations (see \S~\ref{sec:thestar} for further discussion of the spot variability modeling). The eclipses were excluded when training the model, but we did use the model to predict and correct for the systematics and spot-modulations across the eclipses. Finally, we included two change-points in the position component (at $BJD-2454833=2240$ and $2273$, each time the direction of the roll angle variations reverses): the systematics are treated as correlated between each pair of change points, but not across a change-point.

\begin{figure*}
\centering
\includegraphics[width=0.99\textwidth]{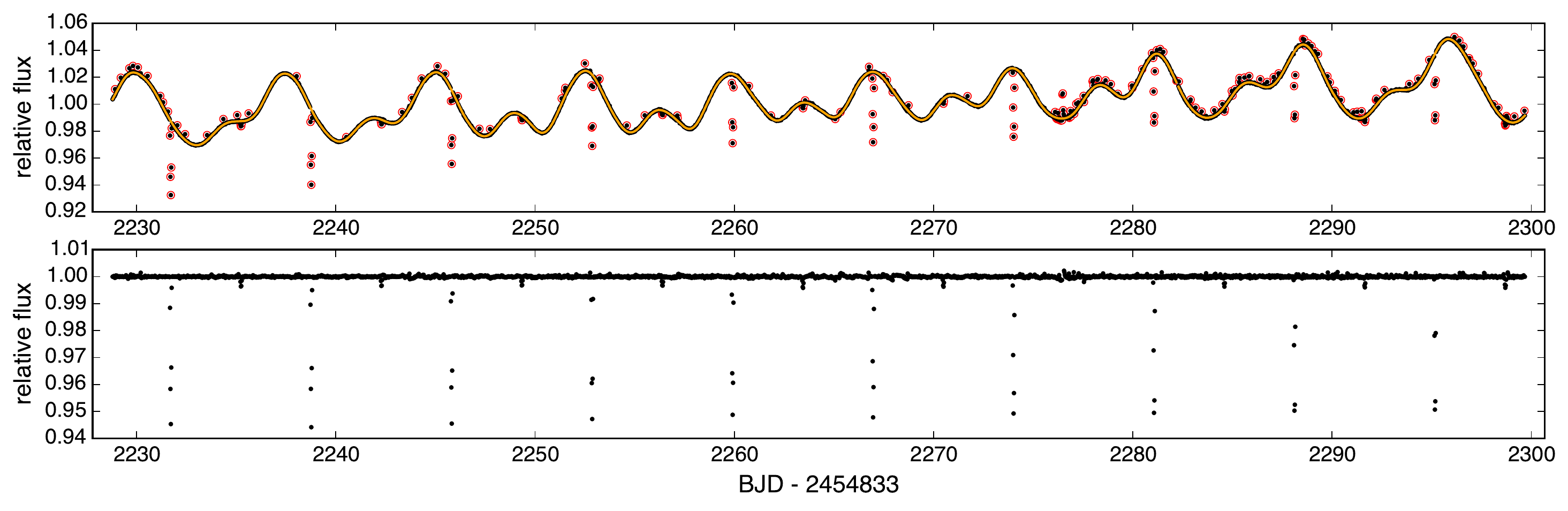}
\caption{Top panel: Systematics corrected $K2$ SAP light curve with our GP stellar variability fit in orange. Observations circled in red were excluded from the variability fit. Bottom panel: Corrected light curve obtained from dividing out the variability fit and excluding outliers.}
\label{fig:lightcurve}
\end{figure*}

We fit for the characteristic amplitude and length scales of the systematics and spot component, the period and evolutionary timescale of the latter, as well as the white noise standard deviation, by maximizing the likelihood subject to log-normal priors on the length scales and log-uniform priors on the other parameters. The priors used were broad enough that they do not affect the fit, merely restricting the model to physically plausible values. Once the covariance parameters are set, we compute, for each cadence, the mean and standard deviation of the predictive distribution of the GP conditioned on the observations. We flag any observations lying more than three standard deviations away from the mean as outliers and repeat the fitting and prediction procedure to ensure the outliers did not affect the fit. The mean of the predictive distribution for the full (systematics + spot) model was subtracted from the data in order to model the eclipses, with the standard deviation of the same predictive distribution serving as our estimate of the photometric errors. These include the white noise term, plus additional errors arising from imperfections in the GP model's ability to reproduce the data. As one would expect, the errors are slightly larger during the eclipses, where the data was not used to constrain the fit. We also evaluated the individual systematic and spot components separately in order to produce a systematics-corrected light curve that preserves astrophysical variability.
The systematics-corrected light curve, variability fit, and final light curve used in eclipse modeling are presented in Figure~\ref{fig:lightcurve}.

\section{HII 2407}
\label{sec:thestar}

The star is a classical member of the Pleiades\footnote{Its categorization in SIMBAD as an RS CVn star is not the correct interpretation of source properties.} with \citet{trumpler1921}, \cite{vanmaanen1945} and \cite{hertzsprung1947} designations; the last is the name by which it is most well known: Hz or HII 2407. The $K2$ identifier is EPIC 211093684. 

A spectral type of G5 is reported in (but not derived by) \cite{mermilliod1992}, which differs from the previous spectral type of K3 presented by \cite{herbig1962}. The star has a $V$ magnitude of 12.19, and $J-K$ color of 0.572. An $R\approx60,000$ spectrum from $\sim3800-8000$ \AA\ of HII 2407 was obtained on UT 9-20-2015 using HIRES \citep{vogt1994} on the Keck I telescope. The estimated spectral type is K1-K1.5 based on the line ratios discussed by \cite{basri1990}. For comparison, a spectral type of K1 corresponds to \teff\ = 5170 K according to the \cite{pm2013} temperature scale.

Independent analysis of an $R\approx20,000$ spectrum over the range of 6450-6850 \AA\ taken with WIYN/Hydra in December 1999 produces an effective temperature estimate \teff=$4970\pm95$ K. This result is based on measurement of 51 lines using the ARES program\footnote{\url{http://www.astro.up.pt/~sousasag/ares/}}, and is consistent with a K2 spectral type from the empirical relations of \cite{pm2013}.

The star is known as a variable, and was identified  by \cite{soderblom1993a,soderblom1993b} to have \ion{Li}{1} 6707 \AA\ absorption as well as weak H$\alpha$ and \ion{Ca}{2} triplet core emission, and as a weak x-ray emitter by \cite{stauffer1994} -- all signs of youth that are consistent with the properties of many other low mass Pleiades members. From the HIRES spectrum, we measure EW(Li) = $43.8\pm5.2$ m\AA.

Color-magnitude diagrams show the star to sit firmly on the main sequence with no photometric excess indicative of multiplicity.  Observations conducted with the Palomar 60" telescope and the Robo-AO instrument \citep{baranec2014} confirm that HII 2407 is an apparently single star at wide separations. No far-red optical companions are detected brighter than the $5\sigma$ contrast limits of 2 mag, 4.25 mag, and 5.5 mag fainter than the star at separations from the star of 0.5", 1.5" and 3.5", respectively.  We use this information below to rule out any ``third light" contamination in the eclipse fitting part of our analysis.

The star was classified as spotted by \cite{norton2007} from analysis of its WASP light curve (designation 1SWASP J034942.26+242746.8), but the eclipses were not identified by those authors. Our reanalysis of the WASP light curve using the WASP transit-search algorithm confirmed the eclipses and yielded the following orbital ephemeris:
\begin{align*}
P &= 7.05046 \pm 0.00003 \ \mathrm{days} \\
\mathrm{HJD}_0 &=  2455302.0983 \pm 0.0019, 
\end{align*}
consistent with our $K2$-derived ephemeris, presented in Table~\ref{table:orbitparams}.

However, HII 2407 was reported as a single-line spectroscopic binary by \cite{mermilliod1992}, who derived a 7.05 day orbital period with zero eccentricity. This period is the same as the 7.05 day eclipse period reported above from the $K2$ analysis.  Below we combine absolute radial velocity measurements from Table 7 of \cite{mermilliod1992} with the $K2$ photometry to fit for the system orbital and stellar parameters. Analysis of the HIRES spectrum revealed no signs of a secondary set of lines brighter than a few percent of the primary at wavelengths shorter than $\sim$8000 \AA, with this limit applicable for radial velocity separations of $>$10 \kms\ between the primary and any putative secondary.

We evaluated the rotation period by modelling the out-of-eclipse light curve using a Gaussian process (GP) model with likelihood:

\begin{equation}
\mathcal{L} = \frac{1}{\sqrt{(2 \pi)^n |K|}} \exp \left( -\frac{1}{2} 
\mathbf{y}^{\rm T} K^{-1} \mathbf{y} \right)
\end{equation}

where $\mathbf{y}$ is a vector of $n$ (normalised) flux measurements, and the elements of the covariance matrix $K$ are given by

\begin{multline}
K_{ij} = k(t_i,t_j) \\
= A^2 \exp \left\{-\Gamma \sin^2\left[\frac{\pi}{P} |t_i-t_j| \right] - \frac{(t_i-t_j)^2}{2L^2} \right\} \\
+ \sigma^2 \delta(t_i-t_j)
\end{multline}

where $A$ is an amplitude, $\Gamma$ an inverse length scale, $P$ a period, $L$ an evolutionary time-scale, and $\sigma$ represents the white noise standard deviation, while $\delta(x)$ is the Kronecker delta function. This covariance function gives rise to a family of functions which display periodic but slowly evolving behaviour, and has previously been used to model the light curve of active stars \citep[e.g.][]{aigrain2012}. The GP model was implemented in {\sc Python} using the {\sc george} package \citep{ambikasaran2014}. To speed up the computation, the light curve was sub-sampled by selecting 500 data points at random. The posterior distribution for $P$ was then evaluated (while marginalizing over the other parameters) using an affine-invariant Markov Chain Monte Carlo (MCMC) implemented in the {\sc emcee} package \citep{foremanmackey2013}. The priors used were uniform in natural log between $-10$ and $10$ for all parameters, and we ran 36 parallel chains of 700 steps each, discarding the first 200 as burn-in. The resulting estimate of the rotation period is $P_{\rm rot}=7.45\pm0.07$.

Our value is inconsistent with both the 7.291 day period previously reported by \cite{hartman2010} from HATNET and the 7.748 day period reported by \cite{norton2007} from a Fourier periodogram analysis of the SuperWASP archive. We note that a separate Lomb-Scargle periodogram analysis of the $K2$ light curve yielded a rotation period of 7.28$\pm$0.30 days (where this approximate uncertainty is estimated from the full-width half-maximum of the oversampled periodogram peak), consistent with the HATNET value. We ultimately adopt the rotation period from the GP modeling in our final analysis.

The $K2$-derived photospheric rotation period can be considered in combination with $v \sin i$ values from the literature. \cite{mermilliod2009} measured $5.2\pm0.9$ \kms\ while \citet{queloz1998} tabulated $6.3\pm0.8$ \kms.  From these two values we calculate a primary radius of $R_1$ = 0.77 $\pm$ 0.13 \rsun\ in the first case or $R_1$ = 0.93 $\pm$ 0.12 \rsun\ in the second. Notably, the radius calculated from the Stefan-Boltzmann law ($\sim$0.72 \rsun) assuming our measured \teff\ and a luminosity from the literature is consistent within error with the smaller radius estimate above, but not the larger value. We note that the large uncertainty in the primary radius is dominated by the $v\sin{i}$ measurement error. 

For comparison, from the evolutionary models of \cite{siess2000} and assumed values of \teff\ = 4764 K and $L$ = 0.29 L$_\odot$, \cite{wright2011} estimated a radius of 0.74 $R_\odot$ and a mass of 0.83 M$_\odot$ for HII 2407.  \cite{hartman2010} reported 0.717 \rsun\ and 0.817 \msun\ from the K-band magnitude and \cite{Y2} isochrones while \cite{bouvier1998a, bouvier1998b} found 0.81 M$_\odot$ from the $I$ magnitude, an assumed age of 120 Myr, and \cite{baraffe1998} models. 

Ultimately, we adopt the following as the final primary parameters: spectral type of K2$\pm$1, $T_\mathrm{eff,1}$=4970$\pm$95 K, $\log{(L_1/L_\odot)}$=-0.54, $M_1$=0.81$\pm$0.08 \msun, and $R_1$=0.77 $\pm$ 0.13 \rsun. Our adopted values for HII 2407 are mostly consistent with the highly precise measurements of well-studied double-lined eclipsing binaries. Among comparable main sequence systems compiled by \cite{torres2010}, K1-K3 types have masses in the range 0.764-0.934 \msun, radii of 0.768-0.906 \rsun, \teff\ in the range 4720-5220 K, and luminosities of \loglsun\ -0.515 to -0.303. We note our mass uncertainty of 10\% is arbitrary and intended to be conservative. For comparison, there is a $\sim$7\% dispersion in the masses of K1-K3 benchmarks discussed above. HII 2407 has a luminosity slightly lower than typical, perhaps owing to the presence of spots given its relatively young age.

\begin{figure}
\centering
\includegraphics[width=0.45\textwidth]{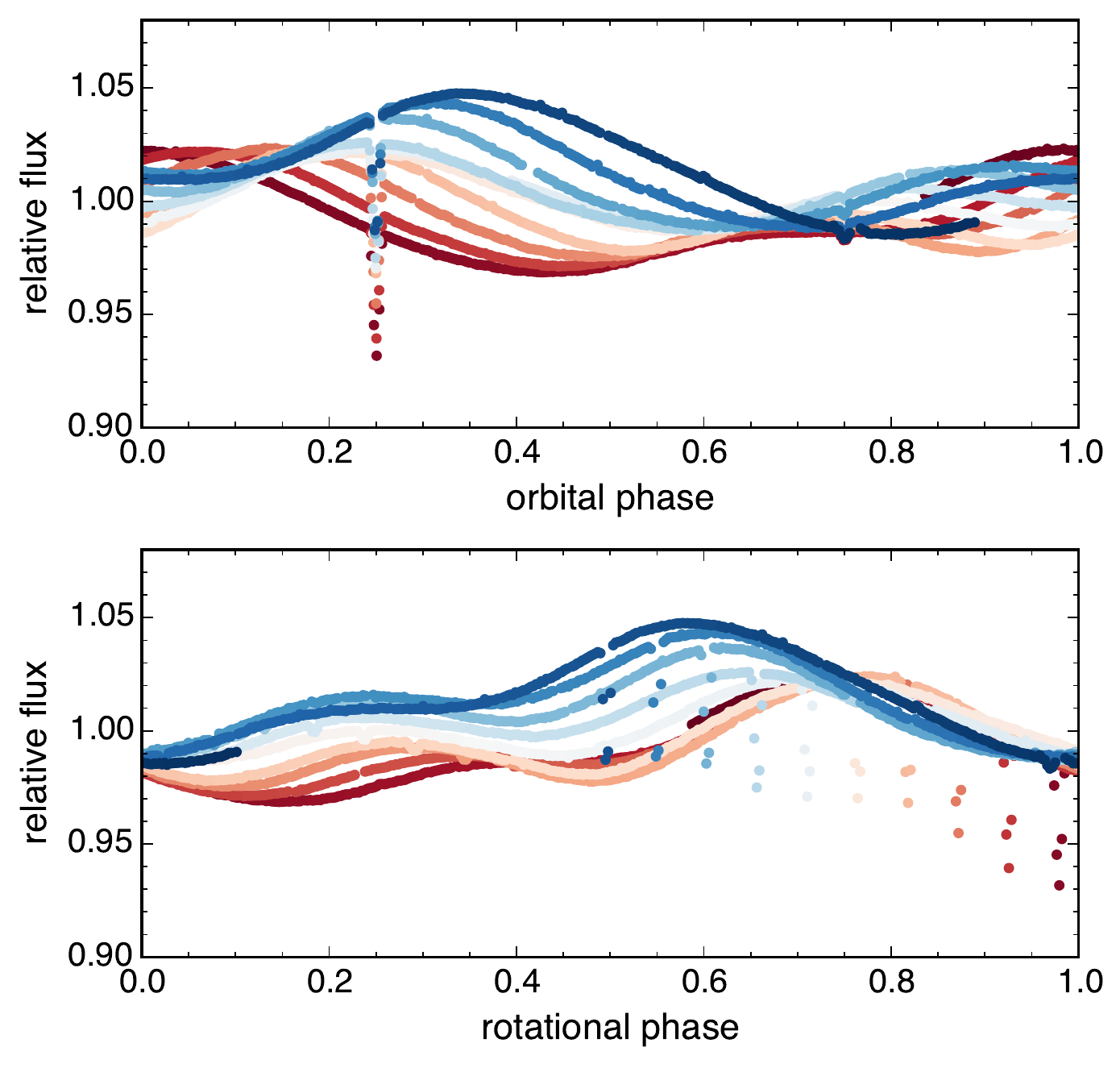}
\caption{Systematics-corrected $K2$ light curve phase-folded on the orbital period (top) and on the rotational period (bottom). Points are colored according to the time of observation. Rotational modulation of starspots is clearly demonstrated. The variable amplitude of the spot signature suggests a changing spot fraction. Though the eclipses are clearly not in phase with the rotational period, we discuss in  \S~\ref{sec:discussion} the likelihood that the system is tidally synchronized with the difference in spot and orbital periods due to a latitudinal gradient in the rotation rate.}
\label{fig:spots}
\end{figure}

\section{Orbital Parameter Fitting}
\label{sec:fitting}

We used the \jktebop\footnote{\url{http://www.astro.keele.ac.uk/jkt/codes/jktebop.html}} orbit-fitting code \citep[][and references therein]{southworth2013} to derive the orbital and stellar parameters for the HII 2407 system. The code is based on the Eclipsing Binary Orbit Program \citep{popper1981, etzel1981}, which relies on the Nelson-Davis-Etzel biaxial ellipsoidal model for well-detached EBs \citep{nd1972, etzel1975}. \jktebop\ models the two components as biaxial spheroids for the calculation of the reflection and ellipsoidal effects, and as spheres for the eclipse shapes. 

Our procedure of removing the out-of-eclipse variability also eliminates gravity darkening, reflected light, and ellipsoidal effects from the light curves. As such, parameters related to these effects are not included in the \jktebop\ modeling. Additionally, out-of-eclipse observations are excluded in order to reduce the effect these observations have on the $\chi^2$ calculation and to expedite the fitting process.  The observational errors were iteratively scaled by \jktebop\ to find a $\chi^2_\mathrm{red}$ close to 1. A single outlier towards the center of secondary eclipse, located more than 3-$\sigma$ above the eclipse minimum, was deemed systematic in nature and was excluded from further analysis.

The integration times of \Kepler\ long cadence data are comparable to the eclipse durations, resulting in ``phase-smearing'' of the light curve. The long exposure times were accounted for in \jktebop\ by numerically integrating the model light curves at ten points in a total time interval of 1766 seconds, corresponding to the \Kepler\ long cadence duration. 

The code finds the best-fit model to a light curve through Levenberg-Marquardt (L-M) optimization. The initial L-M fitting procedure requires reasonable estimates of the orbital parameters to be determined. Period estimates were obtained using Lomb-Scargle \citep{lomb1976, scargle1982} periodogram analysis. Approximations of the ephemeris timebase, $T_0$, were obtained by manually phase-folding the light curves on the periodogram period. 

Holding the period and ephemeris timebase fixed, initial L-M fits are performed in succession for the remaining orbital parameters: the central surface brightness ratio, $J=(T_\mathrm{eff,2}/T_\mathrm{eff,1})^4$ (which can be approximated by the ratio of the eclipse depths for circular orbits), the sum of the relative radii, $(R_1+R_2)/a$, the ratio of the radii, $k=R_2/R_1$, the orbital inclination, $i$, and the quantities $e\cos\omega$ and $e\sin\omega$, where $e$ and $\omega$ are the eccentricity and periastron longitude, respectively. We find an initial estimate for $J$ from the ratio of secondary to primary eclipse depths, which is $\approx 1/13$. For circular orbits this ratio approximately corresponds to a temperature ratio of $\sim$0.53 between the secondary and primary, or a central surface brightness ratio of $\sim$0.08. We find the data to be consistent with a circular orbit, but also explore the possibility of a nonzero eccentricity. Additionally, we incorporate radial velocities (RVs) in the fitting procedure, introducing free parameters corresponding to the RV semi-amplitudes of the primary, $v_{r}$, and the systemic RV, $\gamma$.

After successively increasing the number of free parameters in the fit, a final L-M fit was performed allowing all relevant parameters to be free. In modeling each system, we assumed a linear limb-darkening law for both components and held the limb-darkening coefficients fixed at 0.7, corresponding to the mean value tabulated by \cite{sing2010} for the \Kepler\ bandpass and solar metallicity stellar atmospheres with $3500 \leq$ \teff\ $\leq 5500$, $4.0 \leq \log{g} \leq 4.5$.

We also explored a quadratic limb-darkening law, adopting limb darkening coefficients of $a_1, b_1$ = 0.70, 0.04 for the primary (corresponding to the mean values tabulated by \cite{claret2012} for solar metallicity atmospheres with 4400 K $\leq$ \teff\ $\leq$ 4800 K, $\log{g}$=4.5) and $a_2, b_2$ = 0.41, 0.29 for the secondary (corresponding to the mean values for 3000 K $\leq$ \teff\ $\leq$ 4000 K, $\log{g}$=5.0). Using a quadratic limb-darkening law in this case provided essentially no improvement to the quality of the light curve fit. We suggest that grazing eclipses, spot activity, the quality of the $K2$ photometry, and the light curve processing procedures may all contribute to some degree in making it difficult to constrain limb-darkening parameters for this system.

Robust statistical errors on the best-fit model parameters are then found through repeated Monte Carlo (MC) simulations in which Gaussian white noise commensurate to the observational errors is added to the best-fit model. A new L-M fit is performed on the perturbed best-fit model and the new parameters are saved as links in the MC chain. The final orbital parameters for each system are then given by the original L-M best-fit, with uncertainties given by the standard deviations determined from the MC parameter distributions. 

\begin{figure*}
\centering
\includegraphics[width=0.95\textwidth]{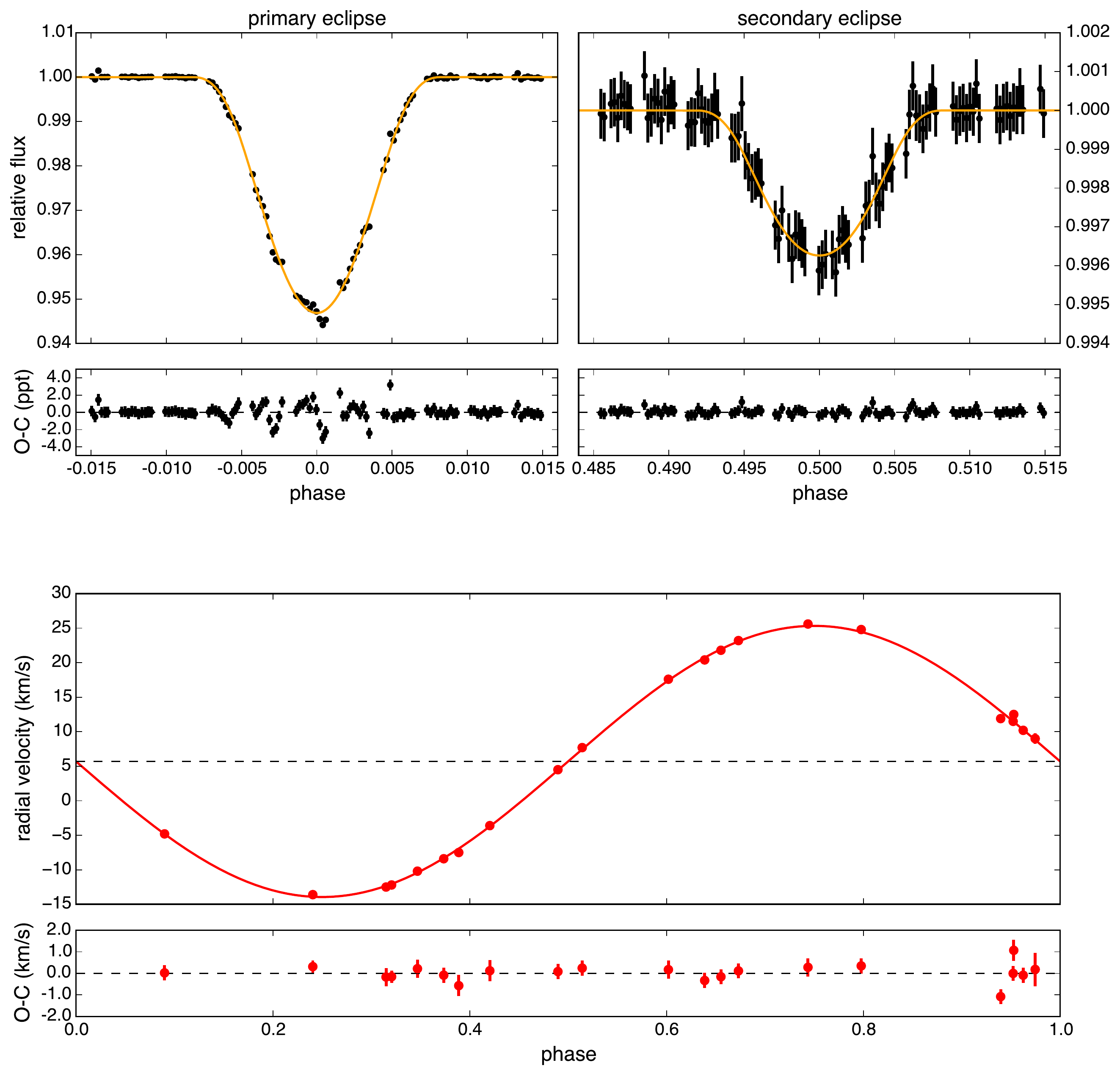}
\caption{Best-fit \jktebop\ model to the $K2$ photometry (top panels) and the \cite{mermilliod1992} radial velocities (bottom panel). For each panel the residuals of the best fit model are plotted below. Measurement uncertainties in the top left and bottom panels are smaller than the points themselves. The increased scatter seen in primary eclipse is potentially due to spot activity and/or artifacts from the \Kepler\ data reduction pipeline. The horizontal dashed line in the bottom panel indicates the best-fit systemic radial velocity.}
\label{fig:fit}
\end{figure*}

The best-fit \jktebop\ model light curve and radial velocity curve are presented in Figure~\ref{fig:fit} with details given in Table~\ref{table:orbitparams}. The $\chi^2_\mathrm{red}$ of the best fit is 1.04 for the light curve with out of eclipse observations removed. We also present in Table~\ref{table:orbitparams} the best-fit parameters in the case of an eccentric orbit (where $e\cos\omega$ and $e\sin\omega$ are allowed free), which are completely consistent with the corresponding parameters in the circular orbit solution. The best-fit eccentricity in this case was $e$=0.0044 $\pm$ 0.0049, consistent with zero. We thus adopt the circular orbit solution for the analysis that follows.

\begin{deluxetable*}{lcrll} 
\tabletypesize{\footnotesize} 
\tablewidth{0.95\textwidth} 
\tablecaption{ Best-fit Orbital Parameters \label{table:orbitparams}} 
\tablehead{ 
\colhead{Parameter} & \colhead{Symbol} & \colhead{Value} & \colhead{1-$\sigma$ Error} & \colhead{Units}
}
\startdata 
Central surface brightness ratio & $J$ & \sbratio  & $\pm$ \esbratio  & \\
Sum of fractional radii & $(R_1+R_2)/a$ & \sumradii & $\pm$ \esumradii & \\
Ratio of radii & $R_2/R_1$ & \ratradii & $\pm$ \eratradii\footnote{The statistical uncertainty in $R_2/R_1$ is not reliable in the absence of a flux ratio measurement due to the intrinsic degeneracies of EB lightcurves, particularly for circular orbits (see \S~\ref{sec:discussion}). The HIRES spectrum provides an upper limit to the flux ratio of $\sim$5\%, which corresponds to an upper limit for the radius ratio of $\sim$0.9 using the \jktebop\ temperature ratio or $\sim$0.6 from the PHOEBE temperature ratio, as discussed in \S~\ref{sec:fitting}.} & \\
Inclination & $i$ & \inclination & $\pm$ \einclination & deg \\
Period & $P$ & \period & $\pm$ \eperiod & days \\
Time of primary minimum & $T_0$ & \tzero & $\pm$ \etzero & BJD \\
Radial velocity amplitude & $v_{r}$ & \rvprim & $\pm$ \ervprim & \kms \\
Systemic radial velocity & $\gamma$ & \rvsys & $\pm$ \ervsys & \kms 
\\
\\
\multicolumn{5}{c}{\emph{Eccentric Orbit Parameters}} \\
\\
Central surface brightness ratio & $J$ & 0.0589 & $\pm$ 0.0031  & \\
Sum of fractional radii & $(R_1+R_2)/a$ & 0.0587 & $\pm$ 0.0018 & \\
Ratio of radii & $R_2/R_1$ & 0.268 & $\pm$ 0.023 & \\
Inclination & $i$ & 87.72 & $\pm$ 0.13 & deg \\
Eccentricity, periastron longitude combination & $e\cos{\omega}$ & 0.00006 & $\pm$ 0.00025 & \\
Eccentricity, periastron longitude combination & $e\sin{\omega}$ & -0.0044 & $\pm$ 0.0068 &  \\
Period & $P$ & 7.0504823 & $\pm$ 0.0000049 & days \\
Time of primary minimum & $T_0$ & 2456916.65778 & $\pm$ 0.00015 & BJD \\
Radial velocity amplitude & $v_{r}$ & 19.60 & $\pm$ 0.13 & \kms \\
Systemic radial velocity & $\gamma$ & 5.707 & $\pm$ 0.084 & \kms

\enddata 
\tablecomments{Orbital parameters determined from a simultaneous fit of the corrected $K2$ light curve and \cite{mermilliod1992} radial velocities. Statistical parameter uncertainties are 1-$\sigma$ errors determined from 10,000 Monte Carlo simulations with \jktebop. Parameters in the eccentric orbit case were determined from 5,000 Monte Carlo simulations. All parameters in the eccentric case are consistent within error with the circular orbit fit.}
\end{deluxetable*}

Notably, the minimum of primary eclipse is poorly fit by the model, primarily due to the three lowest flux observations. These data correspond to the first three eclipse minima, suggestive of an intrinsic variability origin to the outlying points. However, we can not rule out the possibility that the low fluxes are systematic in nature.

A battery of tests were performed to assess how the quality of fit changed with the inclusion/exclusion of these outliers and neighboring points. Keeping the observational errors fixed, the best-fit $\chi^2_\mathrm{red}$ is minimized by excluding the three low flux outliers. Moreover, exclusion of the entire bottom of primary eclipse (defined here as those observations with relative flux values lower than 0.955) leads to a best-fit with parameters more similar to those found when excluding just the three low flux outliers. Finally, a higher $\chi^2_\mathrm{red}$ is found by forcing the fit to pass through the low flux outliers through excluding only the cluster of observations occurring just prior to the primary eclipse minimum in phase.

However, given that we know the primary exhibits significant spot activity with periodicity similar to that of the binary orbit, we consider the removal of these outliers a contrived choice. Furthermore, given the youth of the system, it is likely that the low-mass secondary is also spotted. In such instances, complicated patterns may arise during eclipses with contributions from both the background and foreground stars \citep[e.g.][]{gillen2014}. As such, we choose to include all observations from primary eclipse in our final fit and suggest that the increased scatter is likely due to spots.
We note that excluding these three observations changes the best-fit temperature ratio by $<$1\%, the inclination by $\sim$0.2 deg, the sum of fractional radii by $\sim$4\% (or $<$1.5-$\sigma$), and the ratio of radii by $\sim$12\% (or $<$1.5-$\sigma$).

Independent of the \jktebop\ analysis, we also modeled the light curve and radial velocities with PHOEBE \citep{prsa2005}. Based on the fact that we could not detect the secondary component in the HIRES spectrum, we can place an upper limit on the optical flux ratio of $\sim$5\%.  After creating an initial model in PHOEBE, we ran an MCMC fitting routine using both the SB1 radial velocities and the detrended $K2$ light curve with the following free parameters: mass ratio, semi-major axis, inclination, effective temperature of the secondary component, potentials of both the primary and secondary components, and light and third-light levels.  We set priors on the mass ratio and semi-major axis such that resulting masses would be consistent with the estimated values, but generally left them free to explore the degenerate parameter space.  We then introduced a penalty in the likelihood function to forbid any models that resulted in the secondary contributing more than 5\% of the flux.

 We then derived the values and posteriors of the quantities which can actually be constrained by this system by propagating the values for the MCMC chains and fitting a Gaussian to the resulting distributions. Assuming a circular orbit, the PHOEBE analysis yields the following values: $(R_1+R_2)/a$ = 0.0506 $\pm$ 0.0006, $a_1 \sin{i}$ = 2.69 $\pm$ 0.05 \rsun, $i$ = 88.09 $\pm$ 0.04 deg, and a temperature ratio of $T_\mathrm{eff,2}/T_\mathrm{eff,1}$ = 0.612 $\pm$ 0.005. This temperature ratio, combined with the assumed primary temperature, implies $T_\mathrm{eff,2}$ = 3040 $\pm$ 60 K. These values are close to those found by \jktebop, though there is a $\sim$15\% difference in the sum of fractional radii and a $\sim$20\% discrepancy between the temperature ratios favored by the two different codes. These differences suggest the statistical uncertainties we report in Table~\ref{table:orbitparams} may not reflect the true uncertainties. A possible etiology of this behavior is the reliance of PHOEBE on stellar atmospheres to convert surface brightness to \teff\ at cool temperatures, in contrast to \jktebop\ which does not rely on such models. We consider the results of both modeling efforts in the analysis that follows.

\section{Discussion}
\label{sec:discussion}

Simultaneous fitting of the light curve and primary radial velocities yield an RV semi-amplitude of the primary, systemic RV, and binary mass function that are entirely consistent with the values reported in \cite{mermilliod1992}. We find a binary mass function $f(M_2)$ = 0.005521 $\pm$ 0.000097 \msun, providing an absolute lower limit of $\sim$6 \mjup\ to the mass of the secondary.

Since HII 2407 is an SB1 binary, the radial velocities contain information only about the projected orbit of the primary component (i.e. $a_1\sin{i}$) and fail to provide us information about the separation between the two stars or the mass ratio, as would be the case in an SB2 binary.  Without significant ellipsoidal variations, the light curve can not constrain the mass ratio and, since the eclipses are merely grazing, does not provide a strong constraint on the radius ratio either.  Instead, the light curve contains robust information only about the sum of fractional radii $(R_1 + R_2)/a$, the inclination, and the temperature ratio.  

Nevertheless, auxiliary information about the system allows for coarse characterization of the secondary. The broad-band photometry places HII 2407 close to the Pleiades single star locus, which provides a rough constraint on the mass ratio of $q \lesssim 0.3-0.4$. This upper limit, combined with the lower limit from the precisely measured mass function, places the companion firmly in the $\sim$0.006-0.4 \msun\ mass range. This, of course, assuming the primary mass from photometry, which again is consistent with benchmark K dwarfs. Moreover, the inclination is robustly constrained and given the well-defined range of dynamical masses for K1-K3 type benchmark double-lined EBs, one can use the radial velocity equation \citep{lehmann1894} to obtain a reasonable, and more precise, approximation for the secondary mass.  

As noted in Table~\ref{table:orbitparams}, the upper limit on the flux ratio from the HIRES spectrum can be used to place an upper limit on the radius ratio. However, significantly different limits arise from the different temperature ratios favored by \jktebop\ and PHOEBE. An upper limit of $R_2/R_1<$0.9 is obtained from the \jktebop\ best-fit $J$ value, while PHOEBE favors a higher temperature ratio that implies $R_2/R_1<$0.6.

Based on assumed parameters for the primary of $R_1$ = 0.77$\pm$0.13 \rsun\ and $M_1 = 0.81\pm0.08$ \msun\ (\S~\ref{sec:thestar}), the companion to HII 2407 has the following properties: $R_2 \approx$ 0.21$\pm$0.04 \rsun, $M_2 \approx$ 0.18$\pm$0.02 \msun (given the primary mass, and the best-fit radial velocity semi-amplitude and inclination). For these parameters and an assumed age of 120 Myr, interpolation of \cite[][hereafter BHAC15]{baraffe2015} models predicts temperatures of $T_1 = 4975$ K  and  $T_2 = 3120$ K. The predicted flux ratio of this configuration is thus $F_2/F_1\sim$0.1 at 8000 \AA\ or $\sim$0.3 at 1.55 \micron\ (L. Prato, private communication). Detection of spectral lines from the secondary is likely possible in the infrared.

\jktebop\ modeling suggests a luminosity ratio $L_2/L_1 \approx k^2 J \approx$ 0.004, which is consistent with the HIRES-determined upper limit on the optical flux ratio. For comparison, assuming an age of 120 Myr and $M_1=0.81$ \msun, the luminosity ratio suggests $M_2\approx$0.11 \msun, from interpolation among either BHAC15 or \cite{siess2000} isochrones. We caution that this ratio is strongly dependent on the poorly constrained ratio of radii. The best-fit central surface brightness ratio corresponds to a temperature ratio of $T_\mathrm{eff,2}/T_\mathrm{eff,1}$ = 0.4953 $\pm$ 0.0049. This ratio suggests a secondary temperature of $T_\mathrm{eff,2}$ = 2460 $\pm$ 90, assuming a 3-$\sigma$ error in the surface brightness ratio. We note this temperature is $\sim$500 K cooler than predictions from BHAC15 or \cite{siess2000} models for a star with the assumed secondary mass. The temperature ratio favored by PHOEBE, however, produces a secondary temperature that is in much better agreement with models.

\begin{figure*}
\centering
\includegraphics[width=0.95\textwidth]{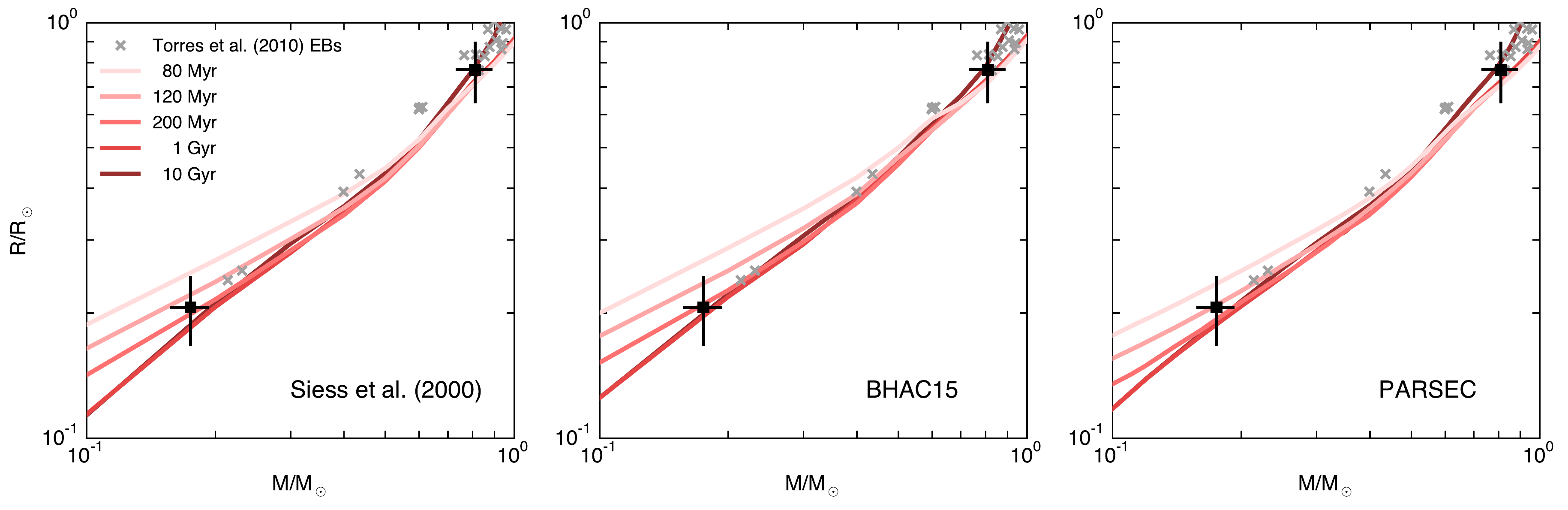}
\caption{Isochrones in the mass-radius plane with the components of HII 2407 and benchmark EBs from \cite{torres2010} overplotted. From left to right, the evolutionary models depicted are from \cite{siess2000}, \cite{baraffe2015}, and \cite{bressan2012}. All models plotted are for solar metallicity (Z=0.02). Unlike the \cite{torres2010} sample, the masses and radii of the HII 2407 components are model-dependent.}
\label{fig:isochrones}
\end{figure*}

The position of the secondary in the mass-radius plane relative to \cite{siess2000}, BHAC15, and PARSEC v1.1 \citep{bressan2012} models (see Figure~\ref{fig:isochrones}) is consistent in each case with an age older than the nominal cluster age of 120 Myr, but within error of the accepted value. The largest discrepancy is present in the BHAC15 models, which imply a significantly older age for the secondary (in other words, the BHAC15 models overpredict the radius at a given mass, if the mass and radius are assumed correct). For the assumed cluster age, the PARSEC models provide the closest match to our estimates of the secondary parameters. However, we note that given the large uncertainties, meaningful constraints on evolutionary models will be obtained only when secondary lines are detected and thus precise masses and radii for both components are measured directly.

Regarding the near-coincidence of the binary orbital and the stellar rotational periods, the $\sim$7 day rotation period of the primary is typical of single Pleiads with masses in the range 0.6-0.8 \msun\ \citep{hartman2010}. The $\sim$0.4 day difference between the rotational and orbital frequencies corresponds to 0.048 radian/day. This is comparable to the equator-to-pole difference in rotational frequency found in Doppler imaging studies of differential rotation in young K dwarfs of similar effective temperature \citep{barnes2005}. In other words, the frequency difference is small enough that if the surface rotation of the primary is locked to the orbit at low latitude, and the spot activity is confined to higher latitudes, the observed frequency difference could arise from surface differential rotation. 

Using equation 4.12 from \cite{zahn1977} for tidal synchronization due to eddy viscosity in a convective star with a tidal Love number of order unity, we obtain a synchronization timescale $\approx3\times10^7$ years for the primary's rotation. While this estimate is somewhat uncertain, it indicates that the synchronization timescale for such a system should exceed the age of the cluster at rotation periods longer than 10 days. This is consistent with the studies of \cite{meibom2006} and \citep{marilli2007}, who found synchronized binaries in clusters of comparable age only at periods less than ten days.
 
We conclude that there is good theoretical and observational support for the interpretation that the similarity between the orbital and photometric periods is causal rather than coincidental, and that the primary's rotation is tidally locked to the orbit.

\section{Summary}

We report the discovery of Pleiades member HII 2407 as an eclipsing binary. The star was known previously as a spectroscopic binary, and we used the literature radial velocities combined with new $K2$ photometry to constrain the fundamental parameters of the system. We revised the spectral type of the primary, provided a new measurement of the rotation period, and demonstrated that the system is likely tidally synchronized. The companion is likely to be a mid-M type, and thus still a contracting pre-main-sequence star given the nominal cluster age. It is the first fundamental calibrator available in this mass and age range. Follow-up infrared spectroscopy, where the flux ratio is more favorable relative to optical spectroscopy, is likely to reveal the lines of the secondary, allowing for dynamically measured masses and elevating the system to benchmark EB status.

\acknowledgments
We thank the referee for suggestions that led to significant improvements in this paper. We thank Lisa Prato for her estimate of the infrared flux ratio and look forward to a direct detection of the secondary.  The material presented herein is based upon work supported in 2015 by the National Science Foundation Graduate Research Fellowship under Grant No. DGE1144469. T.J.D. gratefully acknowledges support from France C\'{o}rdova through the Neugebauer Scholarship. This research was partially supported by an appointment to the NASA Postdoctoral Program at the Ames Research Center, administered by Oak Ridge Associated Universities through a contract with NASA. Support for this work was provided by NASA via grant NNX15AV62G. This paper includes data collected by the Kepler mission. Funding for the Kepler mission is provided by the NASA Science Mission directorate. Some of the data presented in this paper were obtained from the Mikulski Archive for Space Telescopes (MAST). STScI is operated by the Association of Universities for Research in Astronomy, Inc., under NASA contract NAS5-26555. Support for MAST for non-HST data is provided by the NASA Office of Space Science via grant NNX09AF08G and by other grants and contracts. Some of the data presented herein were obtained at the W.M. Keck Observatory, which is operated as a scientific partnership among the California Institute of Technology, the University of California and the National Aeronautics and Space Administration. The Observatory was made possible by the generous financial support of the W.M. Keck Foundation. The authors wish to recognize and acknowledge the very significant cultural role and reverence that the summit of Mauna Kea has always had within the indigenous Hawaiian community.  We are most fortunate to have the opportunity to conduct observations from this mountain. The Robo-AO system was developed by collaborating partner institutions, the California Institute of Technology and the Inter-University Centre for Astronomy and Astrophysics, and supported by the National Science Foundation under Grant Nos. AST-0906060, AST-0960343 and AST-1207891, the Mt. Cuba Astronomical Foundation and by a gift from Samuel Oschin. C.B. acknowledges support from the Alfred P. Sloan Foundation. A.C.C. acknowledges support from STFC grant ST/M001296/1. Funding for WASP comes from consortium universities and from UK’s Science and Technology
Facilities Council.

{\it Facilities:}
\facility{Kepler}, 
\facility{Keck:I (HIRES)}, 
\facility{PO:1.5m (Robo-AO)}, 
\facility{WIYN (HYDRA)}, 
\facility{SuperWASP}.

\end{document}